\documentclass[reprint,showpacs,preprintnumbers,amsmath,amssymb, citeautoscript,prl,superscriptaddress]{revtex4-1}
\usepackage{graphicx,wasysym}
\usepackage{dcolumn}
\usepackage{bm}
\setcitestyle{super}

\usepackage{hyperref}
\usepackage{color}
\usepackage{algorithm}
\usepackage{algpseudocode}

\newcommand{\kB}{k_\mathrm{B}}

\begin{document}
\title{Direct evaluation of rare events in active matter from variational path sampling}

\preprint{APS/123-QED}
\author{Avishek Das}
\thanks{These authors contributed equally}
\affiliation{Department of Chemistry, University of California, Berkeley, CA, 94720, USA \looseness=-1}
\author{Benjamin Kuznets-Speck}
\thanks{These authors contributed equally}
\affiliation{Biophysics Graduate Group, University of California, Berkeley, CA, 94720, USA \looseness=-1}
\author{David T. Limmer}
\email{dlimmer@berkeley.edu}
\affiliation{Department of Chemistry, University of California, Berkeley, CA, 94720, USA \looseness=-1}
\affiliation{Chemical Sciences Division, LBNL, Berkeley, CA, 94720, USA \looseness=-1}
\affiliation{Material Sciences Division, LBNL, Berkeley, CA, 94720, USA \looseness=-1}
\affiliation{Kavli Energy NanoSciences Institute, University of California, Berkeley, CA, 94720, USA \looseness=-1}

\date{\today}

\begin{abstract}
Active matter represents a broad class of systems that evolve far from equilibrium due to the local injection of energy. Like their passive analogues, transformations between distinct metastable states in active matter proceed through rare fluctuations, however their detailed balance violating dynamics renders these events difficult to study. Here, we present a simulation method for evaluating the rate and mechanism of rare events in generic nonequilibrium systems and apply it to study the conformational changes of a passive solute in an active fluid. The method employs a variational optimization of a control force that renders the rare event a typical one, supplying an exact estimate of its rate as a ratio of path partition functions. Using this method we find that increasing activity in the active bath can enhance the rate of conformational switching of the passive solute in a manner consistent with recent bounds from stochastic thermodynamics.
\end{abstract}

\maketitle


\thispagestyle{empty}
\appendix

The constituent agents of active matter-- biomolecules, colloids, or cells-- autonomously consume energy to fuel their motion.\cite{marchetti2013hydrodynamics,shaebani2020computational} 
Their resultant nonequilibrium states have non-Boltzmann phase-space densities and exhibit exotic structural and dynamical collective fluctuations, including motility-induced phase separation and swarming.\cite{cates2015motility,speck2016collective,nemoto2019optimizing,grandpre2021entropy,gompper20202020}  Within these nonequilbrium steady-states, fleeting fluctuations can free particles from external potentials,\cite{woillez2019activated,militaru2021escape,woillez2020nonlocal} nucleate stable phases from metastable ones,\cite{omar2021phase,richard2016nucleation} and assemble passive objects.\cite{stenhammar2015activity, mallory2018active}   
The study of such rare dynamical events within active matter and the calculation of their associated rates is difficult. Traditional equilibrium rate theories like transition state theory and Kramer's theory require knowledge of the form of the steady-state distribution that is not in general available.\cite{nitzan2006chemical} 
Further, only a few numerical methods exist that can be used to tame the exponential computational cost associated with sampling the unlikely fluctuations that lead to transitions between metastable states. Existing methods improve sampling by stratifying or branching stochastic trajectories\cite{allen2009forward,warmflash2007umbrella,cerou2007adaptive} but do not typically employ driving forces to specifically enhance the sampling of these rare events.


Here we present a perspective and an associated numerical algorithm, termed Variational Path Sampling (VPS), for estimating transition rates in active systems using optimized time-dependent driving forces. Our approach relies on a equality between the rate of a rare event in a reference system and a ratio of path partition functions in the reference system and with a driving force that makes the rare event occur with high probability. The VPS algorithm solves a variational problem to approximate the functional form of an optimal time-dependent driving force for this estimate and is applicable to any stochastic dynamics. With VPS we investigate how driven fluids can direct motion into useful function. We apply this technique to study the rate of conformational changes of a passive dimer in a dense bath of active Brownian particles.\cite{fily2012athermal,redner2013structure, bialke2013microscopic} This model exemplifies  how collective active fluctuations around passive solutes can drive self-assembly and speed up transitions between distinct metastable states. \cite{mallory2020universal, sokolov2010swimming} We find the rate to switch between the dimer's two metastable states increases dramatically with increasing activity in the bath, which we rationalize with a recent dissipation bound from stochastic thermodynamics.\cite{kuznets2021dissipation} We study the computational efficiency of rate estimation with VPS and demonstrate its advantage over existing trajectory stratification based methods like Forward Flux Sampling.\cite{allen2009forward}

\begin{figure}[t]
\centering
\includegraphics[width=8.5cm]{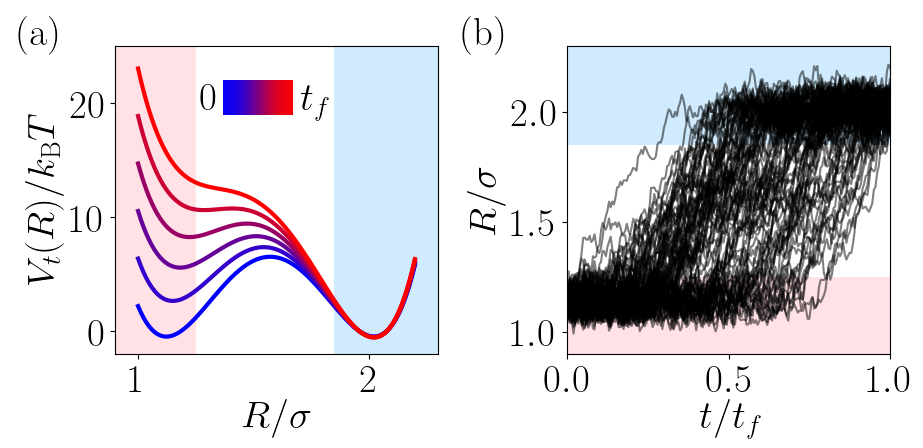}
\caption{Reactive trajectories with VPS. (a)Schematic representation of the total optimal time-dependent potential in an isolated passive dimer as $t$ goes from $0$ to $t_{f}$. Shaded regions are the compact (A, pink) and extended (B, light blue) states. (b) Unbiased reactive trajectories generated with $\boldsymbol{\lambda}(R,t)$.}
\label{fig1}
\end{figure}

We consider a system described by overdamped Brownian dynamics of the form,
\begin{equation}\label{eq:unbiasedlangevin}
    \gamma_{i}\dot{\mathbf{r}}_{i}(t)=\mathbf{F}_{i}[\mathbf{r}^{N}(t)]+\boldsymbol{\eta}_{i}(t)
\end{equation}
where $\dot{\mathbf{r}}_{i}$ is the rate of change of the $i$-th particle's position, $\gamma_{i}$ is the corresponding friction coefficient, and $\mathbf{F}_{i}[\mathbf{r}^{N}(t)]$ is the sum of all conservative, nonconservative and active forces exerted on the $i$-th particle that depends on the full configuration of the $N$-particle system, $\mathbf{r}^{N}$. The final term, $\boldsymbol{\eta}_{i}(t)$, is a Gaussian white-noise with $\langle \eta_{i\alpha}(t)\rangle=0$ and
\begin{equation}\label{eq:langevinnoise}
    \quad\langle \eta_{i\alpha}(t)\eta_{j\beta}(t{'})\rangle=2\gamma_{i}\kB T\delta_{ij}\delta_{\alpha\beta}\delta(t-t{'})
\end{equation}
for component $(\alpha,\beta)$ and $\kB T$ is Boltzmann's constant times the temperature. In order to study the transition rate between  two long-lived metastable states, denoted $A$ and $B$, we define each from a given configuration using the indicator functions, 
\begin{equation}
    h_{X}[\mathbf{r}^{N}(t)] =\begin{cases} 1\quad  \, \mathrm{if} \, \mathbf{r}^{N}(t) \in X \\ 0 \quad\, \mathrm{else}  \end{cases} \, ,
\end{equation}
for either $X=A,B$. In practice this designation requires an order parameter capable of distinguishing configurations and grouping them into these distinct metastable states like that illustrated in Fig.~\ref{fig1}(a) in one dimension. 
Assuming there exists a separation between the time $\tau^\ddag$ required to traverse the transition region between the two metastable states, and the typical waiting time for the transition, the rate $k$ can be evaluated from the probability to observe a transition, per unit time \cite{chandler1978statistical}
\begin{align}\label{eq:sideside}
	 k&=
	  \frac{\langle h_{B}(t_f) h_{A}(0)\rangle}{t_f \langle h_{A}\rangle}
	 =t_f^{-1} \langle h_{B|A}(t_f)\rangle \, ,
\end{align}
where the angular brackets denote an average over trajectories of duration $\tau^\ddag < t_f \ll 1/k$ started from a steady-state distribution in $A$ and $\langle h_{B|A}(t_f) \rangle$ denotes the conditional probability for transitioning between $A$ and $B$ in time $t_f$. When $t_f$ is chosen to satisfy the timescale separation described above, $k$ is independent of time. 

If the transition is rare, most short trajectories are nonreactive leading to difficulties in estimating the rate directly. Instead of trying to evaluate the small transition probability through stratification as other existing methods do,\cite{allen2009forward,warmflash2007umbrella} we instead optimize a time-dependent driving force $\boldsymbol{\lambda}(\mathbf{r}^{N},t)$ that constrains the transition to occur, and evaluate the probability cost associated with adding that force to the original dynamics.  For a general time-dependent force $\boldsymbol{\lambda}$, using the Onsager-Machlup form for the probabilities of stochastic trajectories,\cite{Onsager_Machlup} the rate expression in Eq.~\ref{eq:sideside} can be rewritten as\cite{kuznets2021dissipation}
\begin{align}\label{eq:expestimate}
    k&= 
     t_{f}^{-1} \left  \langle e^{-\Delta U_{\boldsymbol{\lambda}}} \right \rangle_{B|A,\boldsymbol{\lambda}} \, ,
\end{align}
where $\langle \rangle_{B|A,\boldsymbol{\lambda}}$ denotes a conditioned average computed in presence of the additional force. This relation holds for forces $\boldsymbol{\lambda}$ that affect the transition to occur with probability 1, such that the rate in the driven ensemble is $1/t_f$. 
The average is of the exponential of the change in the path action, $\Delta U_{\boldsymbol{\lambda}}$,
\begin{equation}
\Delta U_{\boldsymbol{\lambda}}[\mathbf{X}]=-\int_{0}^{t_{f}} dt\sum_{i} \frac{[\boldsymbol{\lambda}_{i}^{2}-2\boldsymbol{\lambda}_{i}\cdot(\gamma_{i}\dot{\mathbf{r}}_{i}-\mathbf{F}_{i})]}{4\gamma_{i}\kB T} \, ,
\end{equation}
between trajectories generated with the added force and in its absence. The path action and all other stochastic integrals are evaluated in the Ito convention. 

Equation \ref{eq:expestimate} is a direct estimator for a rate employing an auxiliary control system, but it only becomes useful when the protocol $\boldsymbol{\lambda}(\mathbf{r}^{N},t)$ generates trajectories in a manner equivalent to the unbiased reactive trajectory distribution. This is because the expectation can be viewed as an overlap between the two reactive path distributions, and without significant overlap the exponential average is difficult to estimate. We express the optimality of $\boldsymbol{\lambda}$ using Jensen's inequality after taking the logarithm of Eq. \ref{eq:expestimate} to obtain a variational principle,
\begin{equation}\label{eq:variational}
    \ln k\geq -\ln t_{f}-\langle \Delta U_{\boldsymbol{\lambda}}\rangle_{B|A,\boldsymbol{\lambda}} \, .
\end{equation}
If the average change in conditioned path action $\langle\Delta U_{\boldsymbol{\lambda}}\rangle_{B|A,\boldsymbol{\lambda}}$ is minimized over all possible functional forms of $\boldsymbol{\lambda}$, the rate can be obtained directly as a simple ensemble average of $\Delta U_{\boldsymbol{\lambda^*}}$ at the minimizer $\boldsymbol{\lambda}=\boldsymbol{\lambda}^{*}$. 

The optimal control force $\boldsymbol{\lambda}^{*}$ that saturates Eq. \ref{eq:variational} is unique and given by the solution of the backward Kolmogorov equation\cite{Evans_bridge,Chetrite_Touchette_Conditioned,Chetrite_Touchette_control} as detailed in the Supporting Material (SM). Specifically, the optimal force is $2\kB T$ times the gradient of the logarithm of the commitor probability\cite{vanden2010transition} of ending in state $B$ at $t_f$. A schematic illustration of the optimal effective time-dependent potential $V_{t}(R)$ added to a double well potential is illustrated in Fig. \ref{fig1}(a). The resultant force gradually destabilizes the reactant well to ensure the transition almost surely within the short duration $t_{f}$. Viewed in the backwards direction of time, the potential follows the negative logarithm of the relaxation of an initially localized distribution in $B$ to its steady-state. The force is thus optimal in the sense that reactive trajectories, like those in Fig. \ref{fig1}(b), generated with it are drawn from the reference path ensemble with the correct statistical weights. Generically, $\boldsymbol{\lambda}^{*}(\mathbf{r}^N,t)$ is a function of all particle coordinates, so it is not typically tractable to compute. We demonstrate here that one- and two-body representations of $\boldsymbol{\lambda}$ can be sufficiently close to optimal as to estimate the rate accurately even in cases where the rare event is collective, similar to related observations in large-deviation sampling.\cite{ray2018exact,nemoto2016population,jacobson2019direct,das2019variational} 

We study the accuracy and utility of this formalism in a system comprised of an active bath and a passive dimer that can undergo conformational changes between two metastable states. 
 All particles interact pairwise via a Weeks-Chandler-Andersen (WCA) repulsive potential\cite{weeks1971role} 
\begin{equation}
    V_{\mathrm{WCA}}(\mathbf{r})=
    \left\{ 4\epsilon\left[ \left( \frac{\sigma}{r}\right) ^{12}-\left( \frac{\sigma}{r}\right) ^{6}\right] +\epsilon \right\} \Theta(r_{\mathrm{WCA}}-r)
\end{equation}
with energy scale $\epsilon$, and particle diameter $\sigma$, truncated at $r_{\mathrm{WCA}}\equiv 2^{1/6}\sigma$ with the Heaviside function $\Theta$. Active particles experience an additional self-propulsion force of magnitude $v_{0}$, $\mathbf{F}_{i}^{a}(t)=v_{0}\mathbf{e}[\theta_{i}(t)]$
where the director is $\mathbf{e}(\theta_{i})=(\cos\theta_{i},\sin\theta_{i})$ and $\theta_{i}$ obeys $\dot{\theta}_{i}(t)=\xi_{i}(t)$
with,
\begin{equation}\label{eq:thetadiffusion}
    \langle\xi_{i}(t)\rangle=0,\quad\langle\xi_{i}(t)\xi_{j}(t{'})\rangle=2D_{\theta}\delta_{ij}\delta(t-t{'})
\end{equation}
for angular diffusion constant $D_{\theta}$. Passive solutes separated by distance $R$ are bound by a double-well potential
\begin{equation} \label{double_well}
    V_{\mathrm{dw}}(R)=\Delta V\left[ 1-(R-r_{\mathrm{WCA}}-w)^{2}/w^2\right] ^{2} 
\end{equation}
with an energy barrier of height $\Delta V$ between the compact and extended states at $R=r_{\mathrm{WCA}}$ and $R=r_{\mathrm{WCA}}+2w$ respectively.\cite{dellago1999calculation} We study the transition rates between these states, employing indicator functions $h_{A}(t) = \Theta(R_A -R)$ and $h_{B}(t) = \Theta(R-R_B)$ for $R_A = 1.25 \sigma$ and $R_B = 1.85 \sigma$. Conformation transitions like these in dense fluids are collective in origin\cite{dellago1999calculation} and serve as a sensitive probe of the bath.  

The VPS algorithm estimates an optimal force using a low-rank ansatz by iteratively solving the variational problem in Eq. \ref{eq:variational}, and uses this force to directly obtain a rate estimate. 
For computing the rate of isomerization of the passive dimer,  we approximate $\boldsymbol{\lambda}^{*}$ with a time-dependent interaction along the dimer bond vector $\mathbf{R}$, expressed as a sum of Gaussians
\begin{equation}\label{eq:ansatz}
    \boldsymbol{\lambda}(\mathbf{R},t) =\hat{\mathbf{R}}\sum_{p,q=1}^{M_{R},M_{t}}c_{pq}^{(i)} e^{-\frac{(R-\mu_{R,p})^{2}}{2\nu_{R}^{2}}-\frac{(t-\mu_{t,q})^{2}}{2\nu_{t}^{2}}}
\end{equation}
where $c_{pq}^{(1)}=-c_{pq}^{(2)}$ are variational parameters to be tuned, and the locations and widths $\mu_{R,p}$, $\mu_{t,q}$, $\nu_{R}$ and $\nu_{t}$ are held fixed. To impose the conditioning while minimizing $\langle \Delta U_{\boldsymbol{\lambda}}\rangle_{B|A,\boldsymbol{\lambda}}$, we use a Lagrange multiplier $s$ to construct a loss function $\Omega_{\boldsymbol{\lambda}}=\langle \Delta U_{\boldsymbol{\lambda}}\rangle_{\boldsymbol{\lambda}}+s(\langle h_{B|A}\rangle_{\boldsymbol{\lambda}}-1)$. For a general force that does not ensure the transition with unit probability, there is a multiplicative contribution to the estimate of the rate in Eq.~\ref{eq:expestimate} from $\langle h_{B|A}\rangle_{\boldsymbol{\lambda}}$, which for most optimized forces is negligible. 

The optimization problem maps onto the computation of a cumulant generating function for the statistics of the indicator $h_{B}(t_{f})$ studied previously,\cite{Chetrite_Touchette_control,Avishek_Dom_rare} with the short trajectories starting from a steady-state distribution in the initial state. As such we can employ generalizations of recent reinforcement learning procedures to efficiently estimate the gradients of the loss function with respect to the variational parameters.\cite{rose2021reinforcement} Specifically, we modify the Monte-Carlo Value Baseline (MCVB) algorithm\cite{Avishek_Dom_rare} which performs a stochastic gradient descent to optimize $c_{pq}^{(i)}$. We add two preconditioning steps over the MCVB algorithm. First, we generate an initial reactive trajectory using a routine reminiscent of well-tempered metadynamics.\cite{Well_Tempered_Metadynamics} Then we symmetrize the learned force to ensure time translational invariance of the transition paths. We denote this preconditioning algorithm MCVB-T.  Further information is available in the SM.

We first illustrate the systematic convergence of VPS by estimating the isomerization rate of an isolated passive dimer. Such a simplified system allows us to compare to numerically exact results, and study convergence of the force ansatz in the complete basis limit, where $M_{R},M_{t}\to\infty$ and the Gaussians cover the thermally sampled region in $R$ and $t$. For this simple system, we take $\kB T=\gamma = \sigma =\epsilon=1$, $w=0.25\sigma$, with diffusive timescale $\tau=\sigma^{2}\gamma/\kB T$. We simulate the one-dimensional version of Eq. \ref{eq:unbiasedlangevin} along $R$, with $V_{\mathrm{dw}}(R)$ only. For simplicity we define state $A$ by the initial condition $R(0) = r_{\mathrm{WCA}}$, and state $B$ via $R_B = 1.45 \sigma$. To provide a steady-state value in Eq. \ref{eq:sideside}\cite{Avishek_Dom_rare,delarue2017ab}, we use an Euler method and take in this example $t_{f}= \gamma w\sigma /\sqrt{8  \kB T \Delta V}$. We choose $\mu_{R,{p}}$ and $\mu_{t,q}$ evenly distributed in $R/\sigma\in[0.9,1.77]$ and $t \in[0,t_{f}]$, respectively, and $\nu_{R},\nu_{t}$ to be half the distance between Gaussian centers. We consider basis sizes $M_R = M_t = 2-40$, each optimized independently and used to sample $\sim 10^{5}$ transition paths. 

\begin{figure}
\centering
\includegraphics[width=8.5cm]{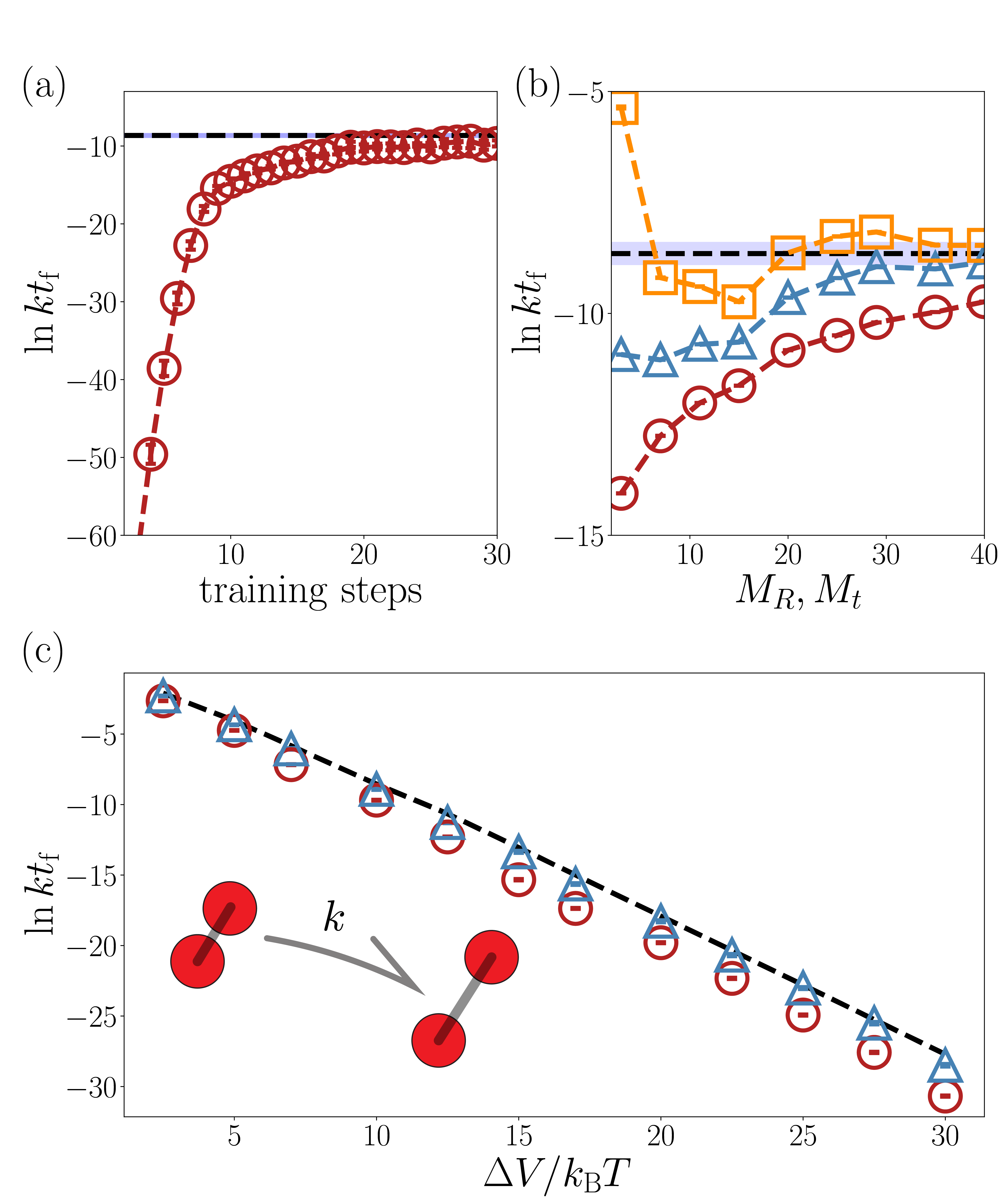}
\caption{Convergence of isomerization rates for an isolated passive dimer. (a) Learning curve for $\Delta V=10\kB T$ and $M_R,M_t=20$. (b) Convergence of the variational rate estimate (circles) and cumulant corrections for $\ell=2$ (triangles) and $\ell=4$ (squares) with basis size as compared to the numerically exact answer (dashed line). (c) Variational (circles) and  $\ell=2$ (triangles) estimate of the rate compared to the exact value (dashed line) with increasing barrier height.}
\label{fig2}
\end{figure}
 Figure \ref{fig2}(a) illustrates a typical learning curve for the control force, showing convergence of the variational rate bound towards the numerically exact rate. The variational estimate requires a basis of $M_R, M_t>40$  to approach the rate to within the statistical uncertainty of the estimate, however alternative estimates with small basis sets can be refined using a cumulant expansion approximation to Eq.~\ref{eq:expestimate}. Specifically, truncating the exact exponential relation at the $\ell$th cumulant as
\begin{equation}\label{eq:cumulant}
    \ln k  \approx -\ln t_{f} + \sum_{n=1}^{\ell} \frac{1}{n!} \frac{d^{n} \ln \left  \langle e^{- \Delta U_{\boldsymbol{\lambda}}} \right \rangle_{B|A,\boldsymbol{\lambda}}}{d \Delta U_{\boldsymbol{\lambda}}^{n}}
\end{equation}
provides an approximation to the rate that converges in the limit that $\ell$ is large. Figure \ref{fig2}(b) illustrates this convergence, where we find that even coarse-representations of the control force can yield close estimates of the rate with only the first few cumulants, illustrating a tradeoff between basis set completeness and statistical efficiency. Sweeping across a wide range of barrier heights in Fig. \ref{fig2}(c), we find excellent agreement between the log-rate from brute force simulations and a truncation of the cumulant expansion to $\ell=2$ using $M_R=80$ and $M_t=30$. 


We next compute the isomerization rate with VPS when the dimer is immersed in an explicit solvent of active Brownian particles with $N=80$ and a total density of $0.6/\sigma^{2}$. The dimer particles have a friction $\gamma_\mathrm{d}=2\gamma$ and the solvent particles have $\gamma_\mathrm{s}=4\gamma$. We take $\gamma=\sigma=\epsilon=1$, $\kB T=0.5$, $\Delta V=7\kB  T$, $\tau=\sigma^{2}\gamma/2\kB T=1$, $D_{\theta}=1/\tau$ and timestep $10^{-5}\tau$. We also change $w=0.45\sigma$ such that the collisional cross-section of the dimer is large. Collisions with active particles transduce energy along the dimer bond and we study the change in the isomerization rate as the bath activity $v_{0}\sigma /\kB T$ is varied from $0$ to $18$. We use a basis size of $M_{R}=M_{t}=50$ distributed between $R/\sigma\in[0.9,2.3]$ and $t\in[0,t_{f}]$ where $t_{f}=0.2\tau$. 
The optimization starts by learning forces $\boldsymbol{\lambda}(\mathbf{R},t)$ for the isolated dimer with WCA interactions between monomers, followed by the MCVB-T algorithm. Then, $\boldsymbol{\lambda}(\mathbf{R},t)$ is optimized in the presence of the bath for $v_{0}=0$ and higher values of $v_{0}$ are initialized from converged forces at the previous $v_{0}$. 

\begin{figure}[t]
\centering
\includegraphics[width=8.5cm]{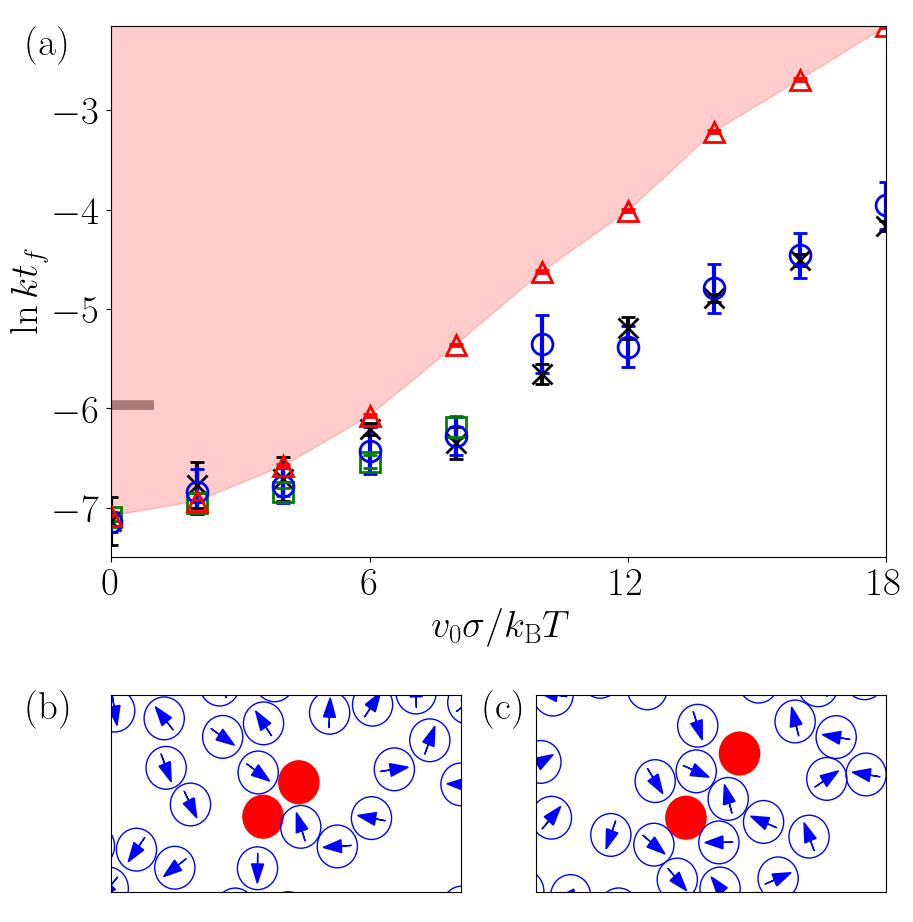}
\caption{Rate enhancement of isomerization in an active fluid. (a) Change in the rate as estimated from direct unbiased simulations (crosses), from exponential estimate (squares), and from histogram reweighting (circles). The excess dissipated heat (triangles) bounds the rate enhancement achievable demarked by the red shaded region. The thick tick mark on the left denotes the rate for the isolated dimer. (b) and (c) Typical snapshots of reactive trajectories of the active bath (blue) and passive dimer (red), at $t=0$ and $t=t_{f}$. 
}
\label{fig3}
\end{figure}

The rate is a strong function of activity, increasing twenty-fold over the range of $v_0$'s considered. While the variational rate estimate from Eq. \ref{eq:variational} is closest for the passive bath, it weakens with increasing $v_{0}$, indicating a growing importance of solvent degrees of freedom in the optimal control force. With converged forces at each $v_{0}$, we run $10^{6}$ trajectories of length $t_{f}$ to compute $k$ from Eq. \ref{eq:expestimate}. This estimate correctly predicts the suppression of $k$ due to passive solvation and can be converged statistically for $v_{0}\sigma/\kB T<9$, which is supported by direct rate estimates from unbiased simulations in Fig. \ref{fig3}(a). Above $v_{0}\sigma/\kB T=9$, the optimized force is not close enough to $\boldsymbol{\lambda}^*$ to estimate $k$ directly through the exponential average or a low order cumulant expansion. 

Provided we have access to the transition path ensemble from direct unbiased simulations or methods like Transition Path Sampling\cite{bolhuis2002transition,ray2018importance, buijsman2020transition} we can supplement the estimate of $k$ using histogram reweighting.\cite{frenkel2019understanding} $k$ satisfies a reweighting relation of the form,
\begin{equation}\label{eq:mbar}
    k=\frac{e^{-\Delta U_{\boldsymbol{\lambda}}}P_{B|A,\boldsymbol{\lambda}}(\Delta U_{\boldsymbol{\lambda}})}{t_{f}P_{B|A,0}(\Delta U_{\boldsymbol{\lambda}})}
\end{equation}
where we have defined $P_{B|A,\boldsymbol{\lambda}}(\Delta U_{\boldsymbol{\lambda}})=\langle \delta(\Delta U_{\boldsymbol{\lambda}}[\mathbf{X}]-\Delta U_{\boldsymbol{\lambda}})\rangle_{B|A,\boldsymbol{\lambda}}$ and similarly for its undriven counterpart $\boldsymbol{\lambda}=0$.
We evaluate $k$ with this estimator by sampling $10^4$ driven and only 6-100 unbiased reactive paths, using the Bennett Acceptance Ratio\cite{shirts2008statistically} to evaluate the ratio of probabilities.  Compared with the brute-force estimate in Fig. \ref{fig3}(a), we find this reweighting predicts $k$ accurately across all values of $v_{0}$ with significantly higher statistical efficiency then a brute force calculation, which validates the accuracy and utility of the control forces. We have compared the VPS rate estimates in the SM, using either Eqs. \ref{eq:expestimate} and \ref{eq:mbar}, to the Rosenbluth variant of Forward Flux Sampling\cite{allen2009forward}, and find that VPS is statistically more efficient and converges more quickly with the number of reactive trajectories.

Access to an ensemble of transition paths in this active system gains us mechanistic insight into the process. The rate enhancement observed for the compact to extended state transition of the passive dimer with bath activity can be understood using recent results from stochastic thermodynamics. Specifically the rate enhancement achievable by coupling a reactive mode to a nonequilibrium driving force is bounded from above by the heat dissipated over the course of the transition.\cite{kuznets2021dissipation} In this case the nonequilibrium driving is afforded by the interactions between the dimer and the active bath, so the bound takes the form
\begin{equation}
\ln k \leq \ln k_0  +\frac{1}{2 \kB T} \langle Q-Q_0 \rangle_{B|A}
\end{equation}
where $k_0$ is the rate at $v_0=0$ and $\langle Q -Q_0\rangle_{B|A}$ is the dissipative heat less its average at $v_0=0$ given by 
\begin{equation}
Q = \int_0^{t_f} dt \sum_{i \in \mathrm{d}} \sum_{j \in \mathrm{s}} \left (\mathbf{\dot{r}}_i-\mathbf{\dot{r}}_j\right ) \cdot \mathbf{F}_\mathrm{WCA}(\mathbf{r}_{ij})
\end{equation}
which is a sum of the total force from the WCA potential of the solvent particles (s) on the dimer (d) times the difference in their velocities in an ensemble at fixed $v_0$ (SM). This bound is verified in Fig. \ref{fig3}(a) for all $v_0$, and saturated at small $v_0$. The specific mechanism of energy transfer from bath to dimer that promotes transitions is clarified by examining reactive trajectories driven by the biasing force and are typical, after removal of the bias from the incomplete basis set. Figures \ref{fig3} (b) and (c) show typical snapshots of the solvated dimer at the start and end of the reaction. Energy transfer results from active particles accumulating around the dimer, and preferentially in its cross-section, pushing it apart into an extended state. This mechanism of action is reminiscent of how nonequilibrium agents 
collect in the corners of mesoscopic gears to power their directed rotation.\cite{mallory2020universal, sokolov2010swimming} 
At low $v_0$, we find the driven isomerization process is efficient, while deviation from the bound at large $v_0$ demonstrates that energy is additionally funneled into non-reactive modes. Further studies showing the unbiased nature of the VPS-sampled transition path ensemble in terms of duration and distribution of transition paths, and quantification of the changing solvation environment with $v_0$ are provided in the SM.

In conclusion, we developed a novel formalism and corresponding algorithm termed Variational Path Sampling to compute rate constants in nonequilibrium systems by optimally driving the systems to transition between metastable states. 
VPS can be used to compute rates in arbitrary stochastic systems and extends the use of optimal control forces in large deviation sampling to transient rare events.\cite{das2019variational,Chetrite_Touchette_control,ray2018exact,dolezal2019large,nemoto2016population} VPS complements trajectory-level importance sampling methods by generating the rare reactive event through a time-series of driving forces instead of a sequence of rare noise histories. We expect this approach to find broad use in rate computations for rare events in dissipative systems throughout the physical sciences and across scales.

{\bf Acknowledgements} AD, BKS and DTL were supported by NSF Grant CHE1954580. The authors thank Dominic Rose, Juan Garrahan and Phillip Geissler for useful discussions.

{\bf Data availability} The source code and data that reproduce the findings of this study are openly available on Zenodo at \href {\doibase 10.5281/zenodo.5763101}{https://doi.org/10.5281/zenodo.5763101}.\cite{das_zenodo}
               
\bibliography{oc_rate.bib}

\end{document}


\beginsupplement
\title{Supplemental Material: \\Direct evaluation of rare events in active matter from variational path sampling}

\preprint{APS/123-QED}
\author{Avishek Das}
\thanks{These authors contributed equally}
\affiliation{Department of Chemistry, University of California, Berkeley, CA, 94720, USA \looseness=-1}
\author{Benjamin Kuznets-Speck}
\thanks{These authors contributed equally}
\affiliation{Biophysics Graduate Group, University of California, Berkeley, CA, 94720, USA \looseness=-1}
\author{David T. Limmer}
\email{dlimmer@berkeley.edu}
\affiliation{Department of Chemistry, University of California, Berkeley, CA, 94720, USA \looseness=-1}
\affiliation{Chemical Sciences Division, LBNL, Berkeley, CA, 94720, USA \looseness=-1}
\affiliation{Material Sciences Division, LBNL, Berkeley, CA, 94720, USA \looseness=-1}
\affiliation{Kavli Energy NanoSciences Institute, University of California, Berkeley, CA, 94720, USA \looseness=-1}

\date{\today}

\maketitle

\thispagestyle{empty}
\onecolumngrid



\setcounter{equation}{0}
\setcounter{figure}{0}
\setcounter{table}{0}
\setcounter{page}{1}
\makeatletter
\renewcommand{\theequation}{S\arabic{equation}}
\renewcommand{\thefigure}{S\arabic{figure}}

\section{Demonstration of optimal control force}
 For ease of notation, we consider states $A$ and $B$ being specific phase space points, $\mathbf{r}^N_A$ and $\mathbf{r}^N_B$, respectively. 
Let $\boldsymbol{\lambda}^{*}(\mathbf{r}^{N},t)=2\kB T\boldsymbol{\nabla}\Phi$, where $\Phi(\mathbf{r}^{N}_{B},t_{f}|\mathbf{r}^{N},t)$ is the log of the probability to end at a single target configuration, $\mathbf{r}^{N}_B$, conditioned on being at $\mathbf{r}^{N}$ at time $t$. This conditioned probability satisfies the logarithmic transform of the backward Kolmogorov equation,\cite{majumdar2015effective,Evans_bridge,Chetrite_Touchette_Conditioned,Chetrite_Touchette_control,carollo2018making} 
\begin{equation}\label{eq:kolmogorov}
    \partial_{t}\Phi+\sum_{i}\biggl[  D_{i}(\boldsymbol{\nabla}_{i}\Phi)^{2}+\boldsymbol{\nabla}_{i}\Phi\cdot\mathbf{F}_{i}/\gamma_{i} + D_{i}\boldsymbol{\nabla}_{i}^{2}\Phi\biggr] = 0
\end{equation}
where $D_{i}=\kB T/\gamma_{i}$ is the diffusion constant and the gradients act on $\mathbf{r}_i$.
The optimal force achieves the reactive transition by construction, rendering $\langle h_{B|A}\rangle_{\boldsymbol{\lambda}^{*}}=1$. The change in path action with the optimal force is
\begin{align}
\Delta U_{\boldsymbol{\lambda}^*}[\mathbf{X}] &=-\int_{0}^{t_{f}} dt\sum_{i} D_i(\boldsymbol{\nabla}_i\Phi)^{2}+ \boldsymbol{\nabla}_i\Phi \cdot \mathbf{F}_{i}/\gamma_{i}-\boldsymbol{\nabla}_i\Phi\cdot \dot{\mathbf{r}}_{i}  
\end{align}
Using Ito's Lemma for the total time derivative of $\Phi$ to eliminate the final term,
\begin{equation}
\Delta U_{\boldsymbol{\lambda}^*}[\mathbf{X}] =-\int_{0}^{t_{f}} dt\left\{ \sum_{i} \left [ D_i(\boldsymbol{\nabla}_i\Phi)^{2}+ \boldsymbol{\nabla}_i\Phi \cdot \mathbf{F}_{i}/\gamma_{i} + D_{i}\boldsymbol{\nabla}_{i}^{2}\Phi \right ] -\dot{\Phi} + \partial_{t}\Phi\right \}
\end{equation}
and substituting the backward Kolmogorov equation
\begin{equation}\label{eq:path-independence}
\Delta U_{\boldsymbol{\lambda}^*}[\mathbf{X}] =\int_{0}^{t_{f}} dt \, \dot{\Phi} = \Phi(\mathbf{r}^{N}_{B},t_{f}|\mathbf{r}^{N}_B,t_{f})-\Phi(\mathbf{r}^{N}_{B},t_{f}|\mathbf{r}^{N}_A,0)
\end{equation}
the change of action can be evaluated exactly. The boundary terms from the exact integration are 0 and the log of the transition probability between $A$ and $B$ in the reference system. The latter can be identified with $\ln \langle h_{B|A}\rangle$, resulting in the equality in Eq.~7 using the definition of the rate $k$ in Eq.~4. This reasoning extends linearly to cases where $A$ and $B$ are collections of configurations. 

\section{Protocol for learning optimal force}

We optimize the time-dependent control force $\boldsymbol{\lambda}(\mathbf{R},t)$ by minimizing $\langle\Delta U_{\boldsymbol{\lambda}}\rangle_{B|A,\boldsymbol{\lambda}}$ over variational parameters $c_{pq}^{(i)}$ using Lagrange multiplier $s$ to impose the $B|A$ conditioning. If $s$ is chosen to be more negative than an approximately estimated threshold value $s^{*}=\ln[\langle h_{B|A}\rangle/(1-\langle h_{B|A}\rangle)]$, the optimized forces drive the reaction with unit probability and $s$ need not be individually optimized. For a rare transition, any choice of $s$ with a magnitude an order or more larger than the energy barrier height will robustly provide forces that always satisfy the $B|A$ conditioning.\cite{Avishek_Dom_rare}

For optimization we use an extension of a reinforcement learning algorithm called Monte Carlo Value Baseline (MCVB).\cite{Avishek_Dom_rare} This algorithm computes the correlation of the gradient of the log of the trajectory probability, called Malliavin weights,\cite{das2019variational,Warren2013} with the instantaneous change in $\Delta U_{\boldsymbol{\lambda}}$ and $h_{B}$ over the course of the trajectory. These yield the exact gradients of the loss-function $\Omega_{\boldsymbol{\lambda}}$ with respect to the tunable parameters, with which a stochastic gradient descent is performed. The MCVB algorithm simultaneously learns the driving force and a corresponding value function, $\mathcal{V}(\mathbf{R},t)=\langle \Delta U_{\boldsymbol{\lambda},t}+sh_{B|A}\rangle_{\boldsymbol{\lambda}|R(t)=R}$, where $\Delta U_{\boldsymbol{\lambda},t}$ contains the integrated action difference only within $[t,t_{f}]$ and the expectation is conditioned on starting from $R$ at time $t$. The value function greatly reduces the  the variance of the gradients at zero cost, allowing better convergence. Our modification to this algorithm, refered to as MCVB-T, is a preconditioning step that enforces time translational symmetry for the log of the bridge probability, $\Phi(\mathbf{r}^{N}_{f},t_{f}|\mathbf{r}^{N},t)=\Phi(\mathbf{r}^{N}_{f},t_{f}-t|\mathbf{r}^{N},0)$, as the reference forces are not explicit functions of time. This is achieved by randomly choosing a $t_{\mathrm{mid}}\in[0,t_{f}]$ for every trajectory used for averaging the force gradient, and applying the force $\boldsymbol{\lambda}(\mathbf{R},t\in[t_{\mathrm{mid}},t_{f}])$ on it only for a duration $[t_{\mathrm{mid}},t_{f}]$. This ensures that trajectories undergoing the transition at late times are accounted for while training the force. 

Details of the MCVB-T algorithm are available in the pseudocode in Algorithm \ref{mcvbt-algo}. The set of Gaussian coefficients parametrizing the force and the value function are denoted in short by $\chi$ and $\psi$ respectively. The MCVB algorithm is a special case of MCVB-T with fixed $t_{\mathrm{mid}}=0$.

\begin{figure}
\begin{algorithm}[H]
	\caption{Monte-Carlo Value Baseline with Time-translation invariance (MCVB-T)}\label{mcvbt-algo}
	\begin{algorithmic}[1]
		\State \textbf{inputs} Gaussian coefficients for a general force  $\boldsymbol{\lambda}_{\chi}(\mathbf{r}^{N},t)$ and value function $\mathcal{V}_{\psi}(\mathbf{r}^{N},t)$
		\State \textbf{parameters} learning rates $\alpha^\chi$, $\alpha^\psi$; total optimization steps $I$; trajectory length $t_{f}$ consisting of $J$ timesteps of duration $\Delta t$ each; number of trajectories $N$
		\State \textbf{initialize} choose initial weights $\chi$ and $\psi$, define iteration variables $i$ and $j$, force and value function gradients $\delta_{P}$, $\delta_{V}$, define functional form for stepwise increments (rewards) $\xi$ to the loss-function $\Delta U_{\boldsymbol{\lambda}}+sh_{B|A}$
		\State $i\gets0$
		\Repeat
		\State Generate trajectories $[\mathbf{X}(t)]$ with first-order Euler propagation starting from uncorrelated steady-state configurations in state $A$. Every trajectory starts experiencing the force $\boldsymbol{\lambda}$ from a random time $t_{\mathrm{mid}}$ which is sampled uniformly from $[0,t_{f}]$. Configurations, times, noises (with variance $2\boldsymbol{\gamma}\kB T\Delta t$), Malliavin weights, integral of value function gradients, and rewards are denoted by $\mathbf{r}^{N}_{j},t_{j},\boldsymbol{\eta}_{j},y_{\chi}(t_{j}),z_{\psi}(t_{j})$ and $\xi(t_{j})=\xi_{j}$ respectively.
		\State $j\gets0$
		\State $\delta_P\gets0$
		\State $\delta_V\gets0$
		\State $y_{\chi}(t_{0})\gets0$
		\State $z_{\psi}(t_{0})\gets0$
		\Repeat
		\State $\dot{y}_{\chi}(t_{j})\gets\boldsymbol{\eta}_{j}\cdot\nabla_{\chi}\boldsymbol{\lambda}_{\chi}(\mathbf{r}^{N}_{j},t_{j})/2\kB T\Delta t$
		\State $y_{\chi}(t_{j+1})\gets y_{\chi}(t_{j})+\Delta t\dot{y}_{\chi}(t_{j})$
		\State $\dot{z}_{\psi}(t_{j})\gets\nabla_{\psi}\mathcal{V}_{\psi}(\mathbf{r}^{N}_{j},t_{j})$
		\State $z_{\psi}(t_{j+1})\gets z_{\psi}(t_{j})+\Delta t\dot{z}_{\psi}(t_{j})$
		\State $\delta_P\gets\delta_P+\xi_{j}y_{\chi}(t_{j+1})-\mathcal{V}_{\psi}(\mathbf{r}^{N}_{j},t_{j})\dot{y}_{\chi}(t_{j}))$
		\State		$\delta_{V}\gets\delta_{V}+\xi_{j}z_{\psi}(t_{j+1})-\mathcal{V}_{\psi}(\mathbf{r}^{N}_{j},t_{j})\dot{z}_{\psi}(t_{j})$
		\State $j\gets j+1$
		\Until{$j=J$}
		\State average $\delta_{P}$,$\delta_{V}$ over $N$ trajectories to get $\overline{\delta}_{P}$, $\overline{\delta}_{V}$
		\State $\chi\gets\chi+\alpha^\chi\overline{\delta}_P$
		\State $\psi\gets\psi+\alpha^{\psi}\overline{\delta}_{V}$
		\State $i\gets i+1$
		\Until{$i=I$}
	\end{algorithmic}
\end{algorithm}
\end{figure}

\begin{figure}[h]
\centering
\includegraphics[width=0.9\linewidth]{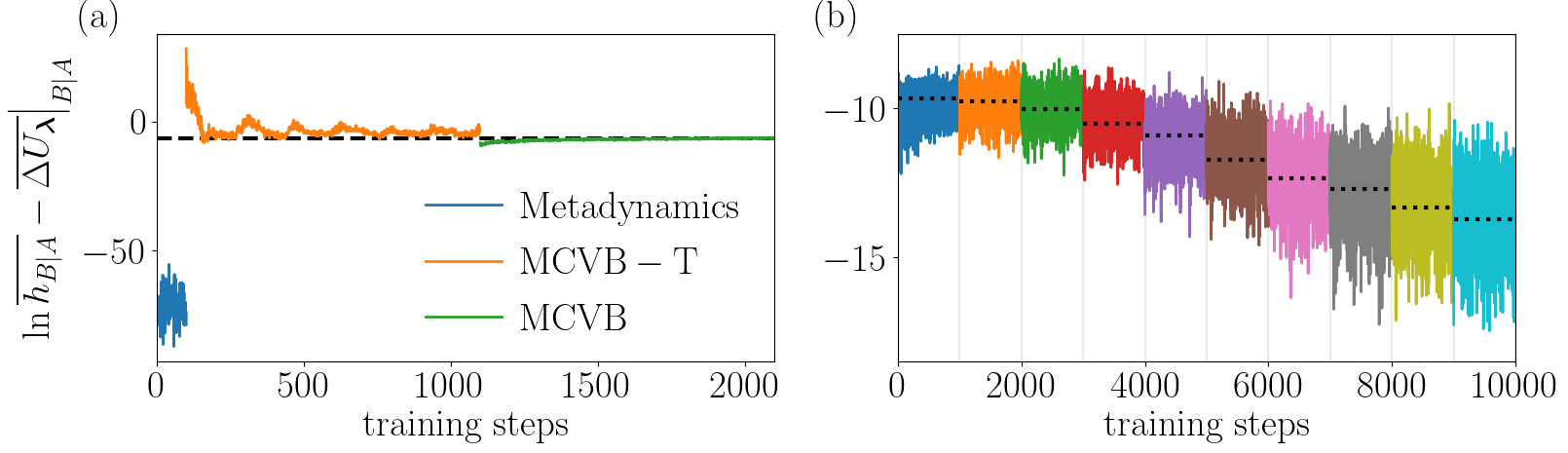}
\caption{Learning curves for variational bound. (a) Optimization for the isolated dimer with 100 steps of well-tempered metadynamics(blue), 1000 steps MCVB-T(orange) and 1000 steps MCVB(green). Black dashed line is $\ln kt_{f}$ for the isolated dimer. (b) Learning curves for 1000 steps each, in presence of the explicit bath with $v_{0}\sigma/\kB T=0,2,4,...,18$. Black dotted lines denote the corresponding converged values.}
\label{sm1}
\end{figure}

Figure \ref{sm1} illustrates all learning curves that led to the results in Fig. 3, plotting $\ln\overline{ h_{B|A}}-\overline{\Delta U_{\boldsymbol{\lambda}}}\bigr|_{B|A}$ as a function of training steps. The averages of this estimator and the gradients are computed over 40 trajectories simulated at each training step. The trajectories are initialized from a collection of steady-state configurations in state $A$. We first learn optimal forces in the absence of the explicit bath in Fig. \ref{sm1}(a), and then optimize these forces in the presence of the bath in Fig. \ref{sm1}(b). We start our optimization by first finding an arbitrary force that ensures the transition with a nonzero probability. We learn initial parameters $c_{pq}^{(i)}$ with a routine similar to well-tempered metadynamics.\cite{Well_Tempered_Metadynamics} Starting with $c_{pq}^{(i)}=0$, at fixed frequency we add $c_{pq}^{(i)}\mapsto c_{pq}^{(i)}+\tau_{m}\omega T_{m}/[T_{m}+\omega\mathcal{N}(R,t)]$, where $\mathcal{N}(R,t)$ is a running histogram of order parameter $R$ up to the current time $t$, and hyperparameters $T_{m}$, $\omega$ and $\tau_{m}$ determine how quickly the force landscape is filled. The blue learning curve in Fig. \ref{sm1}(a) refers to 100 steps of metadynamics run with $\tau_{m}=10t/M_{t}$, $\omega=4000$ and $T_{m}=9000$. We find that the force solely from metadynamics is highly suboptimal compared to the rate bound, indicated by the black dashed line. Starting with a $\boldsymbol{\lambda}$ averaged over all metadynamics steps and with $\mathcal{V}=0$, we next optimize both sets of parameters with MCVB-T and then MCVB each over 1000 steps with learning rates $\alpha^{\chi}=40,\alpha^{\psi}=200$ and $s=-100$. We find that the variational estimate converges tightly to the exact rate bound.

Next we use the converged $\boldsymbol{\lambda}$ and $\mathcal{V}$ to start the optimization in presence of the bath, as illustrated in Figure \ref{sm1}(b). We successively optimize for each $v_{0}\sigma/\kB T\in\{0,2,4,...,18\}$ starting from the previous converged result, each over 1000 steps. Each time we choose $(\alpha^{\chi},\alpha^{\psi})=(0,200)$ for the first 200 steps and $(40,200)$ for the remaining 800 steps. Learning the value function before starting to change the force in this way avoids a brief period of divergence at the beginning of each optimization run.\cite{Avishek_Dom_rare} The results are robust towards changing the learning rates as long as $\alpha^{\psi}$ is kept about 5-10 times of $\alpha^{\chi}$, such that the value function is always approximately accurate whenever the force is being changed.

Results in Fig. 2 were also obtained similar to this protocol, but with no value function. For Fig. 2(c), the initialization parameters $\tau_m$, $\omega$ and $T_m$ are chosen at each barrier height so that at least half of the transitions are reactive.   

\section{Unbiased reactive events from VPS}
\begin{figure}[h]
\centering
\includegraphics[width=1.0\linewidth]{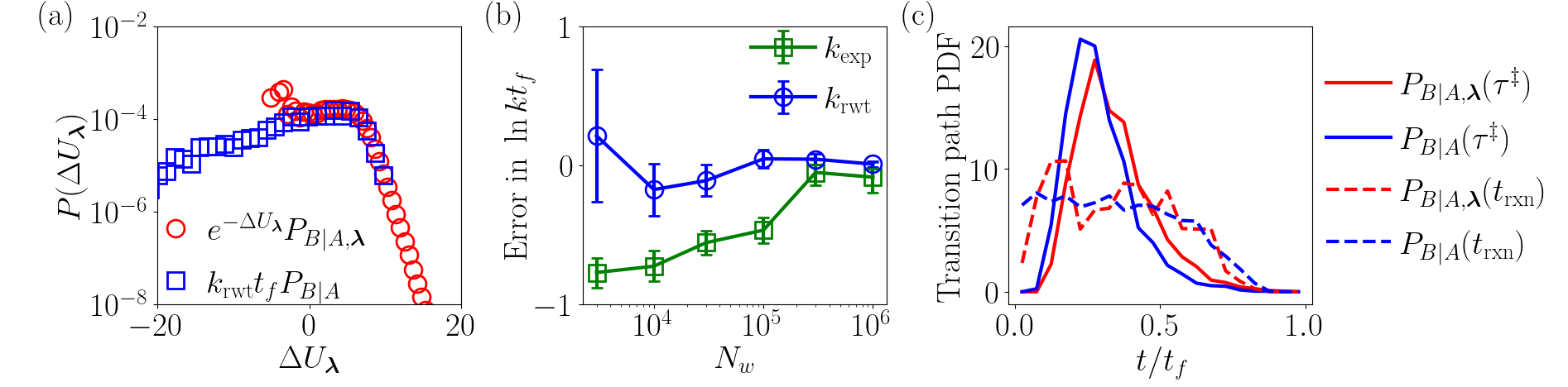}
\caption{Unbiased rates and transition path ensemble with VPS for $v_{0}\sigma/\kB T=8$}. (a) Overlap of the incomplete tilted biased and the unbiased distributions, with the scaling coefficient computed from Bennett Acceptance Ratio. (b) Errors in the rate estimate from Equations 5 ($k_{\mathrm{exp}}$) and 13 ($k_{\mathrm{rwt}}$) as the amount of uncorrelated trajectories used is varied. (c) Probability Density Functions (PDF) of transition path times and reactive escape times in the transition path ensemble, computed from the driven trajectories and unbiased reactive trajectories.
\label{sm2}
\end{figure}

We use Eq. 5 and 13 in the main text to obtain rate estimates from direct simulations using the low-rank optimized force $\boldsymbol{\lambda}$. In Figure \ref{sm2}(a) we have illustrated overlap of the driven distribution $P_{B|A,\boldsymbol{\lambda}}(\Delta U_{\boldsymbol{\lambda}})$ with the unbiased distribution $P_{B|A}(\Delta U_{\boldsymbol{\lambda}})$ after tilting to correct the systematic error. The scaling constant $k_{\mathrm{rwt}}$, which is our estimate for the rate $k$, has been evaluated from Bennett Acceptance Ratio method\cite{shirts2008statistically} by using the tilting exponent as the reduced potential. This overlap is observable only when the driving force $\boldsymbol{\lambda}$ is near-optimal. If the tilted distribution does not contain enough statistics to represent the unbiased distribution, the estimate $k_{\mathrm{exp}}$ from Eq. 5 given by the area under the tilted distribution will underestimate the rate. If the basis set is complete and the exact optimal force $\boldsymbol{\lambda}^{*}$ can be obtained, $\Delta U_{\boldsymbol{\lambda}^{*}}$ will follow a Dirac delta distribution $P_{B|A,\boldsymbol{\lambda}^{*}}(\Delta U_{\boldsymbol{\lambda}^{*}})=\delta(\Delta U_{\boldsymbol{\lambda}^{*}}+\ln kt_{f})$, and the first cumulant will be sufficient to describe the log of the average of the exponential. This is also evident in Eq. \ref{eq:path-independence} where $\boldsymbol{\lambda}=\boldsymbol{\lambda}^{*}$ makes $\Delta U_{\boldsymbol{\lambda}}[\mathbf{X}]$ trajectory independent. In that case, all three estimates of $k$ from Eqs. 5, 7 and 13 will be equal and the unbiased reactive events will be entirely force-assisted rather than being driven by thermal fluctuations.

Figure \ref{sm2}(b) shows the systematic and statistical errors in $\ln kt_{f}$ calculated as $k_{\mathrm{exp}}$ and $k_{\mathrm{rwt}}$ compared to direct unbiased simulation, as the number $N_{w}$ of uncorrelated trajectories of duration $t_{f}$ is varied. At small $N_{w}$, $k_{\mathrm{exp}}$ systematically underestimates the rate due to the full area under $e^{-\Delta U_{\boldsymbol{\lambda}}}P_{B|A}(\Delta U_{\boldsymbol{\lambda}})$ not being accessible because of incomplete overlap, making $k_{\mathrm{exp}}$ formally unbiased but statistically biased. This error disappears with large $N_{w}$. However, the full rate can still be successfully obtained by comparing segments of incomplete distributions. Thus $k_{\mathrm{rwt}}$ incurs much less error with smaller $N_{w}$ and provides a rate estimate that is both formally and statistically unbiased.

Figure \ref{sm2}(c) demonstrates convergence of the transition path ensemble obtained from direct simulations with the optimized forces even before the tilting correction. $P_{B|A,\boldsymbol{\lambda}}(\tau^{\ddag})$ and $P_{B|A}(\tau^{\ddag})$ are distributions of the transition path time $\tau^{\ddag}$ measured as the time after leaving state $A$ and before reaching state $B$ without returning to $A$. $P_{B|A}(\tau^{\ddag})$ is from 2000 reactive trajectories obtained from $10^{6}$ total unbiased simulated trajectories, while $P_{B|A,\boldsymbol{\lambda}}(\tau^{\ddag})$ is from a total of 2000 driven trajectories all of which were reactive. We find convergence in the distribution of transition path times signifying direct access to the nearly unbiased transition path ensemble by using the optimal force. Similarly we compare the distribution of the start time of the reaction $t_{\mathrm{rxn}}\in[0,t_{f}]$ measured as the time the trajectory last leaves $A$ before arriving in $B$. We again find convergence in the driven ensemble compared to the unbiased reactive ensemble indicating the forces $\boldsymbol{\lambda}(\mathbf{R},t)$ are near-optimal at all values of t.

The directly evaluated rates without additional forces used to compare VPS estimates have in most cases been computed from 5 trajectories, each of duration $10^{4}\tau$ with $\tau$ being the diffusive timescale. We compute $k$ using Eq. 4 by evaluating the expectation with a rolling window over the trajectories after relaxing to a steady-state. We deviate from this protocol only for Figure 2(c), where the barrier heights are too large to estimate the rate from direct simulations. Here we use a numerically exact escape rate obtained from Kramer's theory.\cite{zwanzig2001nonequilibrium}
	
\newcommand{\fwcazero}{F^0_\mathrm{wca}}
\newcommand{\fwca}{F_\mathrm{wca}}
\newcommand{\dr}{\dot{\bm{r}}}
\newcommand{\ebf}{ \bm \eta}
\newcommand{\fdw}{ \mathbf{F}_{i}^\mathrm{dw}}
\newcommand{\fdwca}{   \mathbf{F}_{i}^\mathrm{WCA}}
\newcommand{\fds}{ \mathbf{F}_{i}^\mathrm{ds}}
\newcommand{\fsd}{ \mathbf{F}_{j}^\mathrm{sd}}
\newcommand{\fss}{ \mathbf{F}_{j}^\mathrm{WCA}}
\newcommand{\fBl}{ \mathbf{F}_{j}^\mathrm{a}}
\newcommand{\vb}{ \mathbf{F}_{j}^\mathrm{a}}
\newcommand{\ehat}{\mathbf{e}(\theta_{i}(t))}

\newcommand{\uia}{U_\mathrm{ia}}
\newcommand{\unp}{U_\mathrm{np}}
\newcommand{\uip}{U_\mathrm{ip}}
\newcommand{\una}{U_\mathrm{na}}

\newcommand{\kia}{k_\mathrm{ia}}
\newcommand{\knp}{k_\mathrm{np}}
\newcommand{\kip}{k_\mathrm{ip}}
\newcommand{\kna}{k_\mathrm{na}}

\section{Dissipative rate bounds}

Stochastic thermodynamics provides a fundamental speed-limit on the enhancement achievable of the rate $k$ of a rare nonequilibrium transition over a reference equilibrium rate $k_0$ in terms of the excess heat dissipation in the reactive path ensemble,\cite{kuznets2021dissipation} 
\begin{equation} 
    \label{rab}
    \ln k  
     \leq  \ln  k_0+\frac{1}{2 \kB T }\langle Q \rangle_{B|A}
\end{equation}
where $\langle Q \rangle_{B|A}$ is defined as the time-reversal asymmetric contribution from the change in path action between the equilibrium reference and the nonequilibrium system in which it is measured. This bound holds under mild assumptions of instantonic or diffusive transitions and follows from a similar change of measure as leads to Eq. 7, with the additional observation of the time-reversal symmetric contribution of the change in path action being negligible near equilibrium.
Here we show how to arrive at Eq. 14 employing this bound. As in the main text, we assume a separation of timescales between local relaxation and typical transitions so that the rate problem is well-posed.

We consider the rate enhancement afforded by coupling the dimer to an active solvent over the equilibrium passive bath isomerization rate. However, if we simply compute the excess heat dissipated as resulting from the time reversal asymmetric change in path action in turning $v_0$ from 0 to some finite value, the heat will be extensive in the number of solvent degrees of freedom and thus not have a well-defined thermodynamic limit. To mitigate this, we note that the isomerization rate of the dimer would be independent of $v_0$ if the dimer and solvent did not interact. Denoting  $k^{\mathrm{ni}}$ and $k^{\mathrm{ni}}_0$ the rates of isomerization when the dimer is uncoupled to the solvent at finite or zero $v_0$, respectively, then $k^{\mathrm{ni}}=k^{\mathrm{ni}}_0$ and 
\begin{equation}
\ln \frac{k}{k_0} = \ln \frac{k}{k^{\mathrm{ni}}}\frac{k^{\mathrm{ni}}_0}{k_0} \leq \frac{1}{2 \kB T} \left (\langle Q \rangle_{B|A} -\langle Q \rangle_{B|A,0} \right )
\end{equation}
where $\langle Q \rangle_{B|A}$ is the excess dissipation resulting from turning on interactions between the dimer and solvent at finite $v_0$, and $\langle Q \rangle_{B|A,0}=Q_0$ results from turning on interactions between the dimer and solvent at $v_0=0$. The inequality is preserved even though a difference of heats is taken since the second ratio of rates $k^{\mathrm{ni}}_0/k_0$ are both evaluated at equilibrium and thus the symmetric part of the action is zero. This second heat subtracts out the dissipation that is uncorrelated with the isomerization, and the remaining excess dissipation is left finite even when the number of solvent particles is large, so long as the dimer is correlated with the solvent over a finite distance.

The full path action for a system at finite $v_0$ in the presence of dimer-solvent interactions is 
%
%
\begin{align}
U_{v_0}  = - \frac{1}{4 k_B T} \int_0^{t_\mathrm{f}} dt \, &\gamma^{-1}_\mathrm{d} \sum_{i\in\mathrm{d}} \left [ \gamma_\mathrm{d} \dot{r}_i+ \nabla_i V_{\mathrm{dw}} 
-\sum_{j \in \mathrm{d}} \mathbf{F}_{\mathrm{WCA}}(\mathbf{r}_{ij})-\sum_{j \in \mathrm{s}} \mathbf{F}_{\mathrm{WCA}}(\mathbf{r}_{ij}) \right ]^2 \nonumber \\
&+\gamma^{-1}_\mathrm{s} \sum_{i\in\mathrm{s}} \left [ \gamma_\mathrm{s} \dot{r}_i - v_0 \mathbf{e}[\theta_i] 
-\sum_{j \in \mathrm{d}} \mathbf{F}_{\mathrm{WCA}}(\mathbf{r}_{ij})-\sum_{j \in \mathrm{s}} \mathbf{F}_{\mathrm{WCA}}(\mathbf{r}_{ij}) \right ]^2 - \frac{1}{4 D_\theta} \int_0^{t_\mathrm{f}} dt \sum_{i \in \mathrm{s}} \dot \theta_i^2 
\end{align}
and using the convention that $v_0$ is invariant under time-reversal,\cite{grandpre2021entropy} the dissipated heat associated with turning on interactions between the solvent and dimer is
%
\begin{equation} \label{Qint}
Q(\tf) =  \int_0^{\tf} dt \left[  \sum_{i\in \mathrm{d}}  \sum_{j\in \mathrm{d}} \dr_i \mathbf{F}_{\mathrm{WCA}}(\mathbf{r}_{ij}) +  \sum_{j\in \mathrm{s}}\sum_{i\in \mathrm{d}} \dr_j \mathbf{F}_{\mathrm{WCA}}(\mathbf{r}_{ji})  \right]
\end{equation}
and is the same if $v_0=0$ or is nonzero. Since $\mathbf{F}_{\mathrm{WCA}}(\mathbf{r}_{ji})=-\mathbf{F}_{\mathrm{WCA}}(\mathbf{r}_{ij})$ we find the definition of the $Q$ in Eq. 15.

We measure  $ \langle Q \rangle_{B|A}(t_f)$ by averaging Eq. \ref{Qint} over reactive trajectories of length $t_f = .2\tau$ sampled from long, $2 \times 10^8 $ time-step, simulations in the nonequilibrium steady state at fixed $v_0$, with all other parameters as in the main text. Assuming transitions are uncorrelated, we compile $Q$ samples from $24-96$ independent simulations, and use this data to calculate a mean and standard error, as depicted in Fig. 3 (red triangles). 

\begin{figure}[h]
\centering
\includegraphics[width=0.6\linewidth]{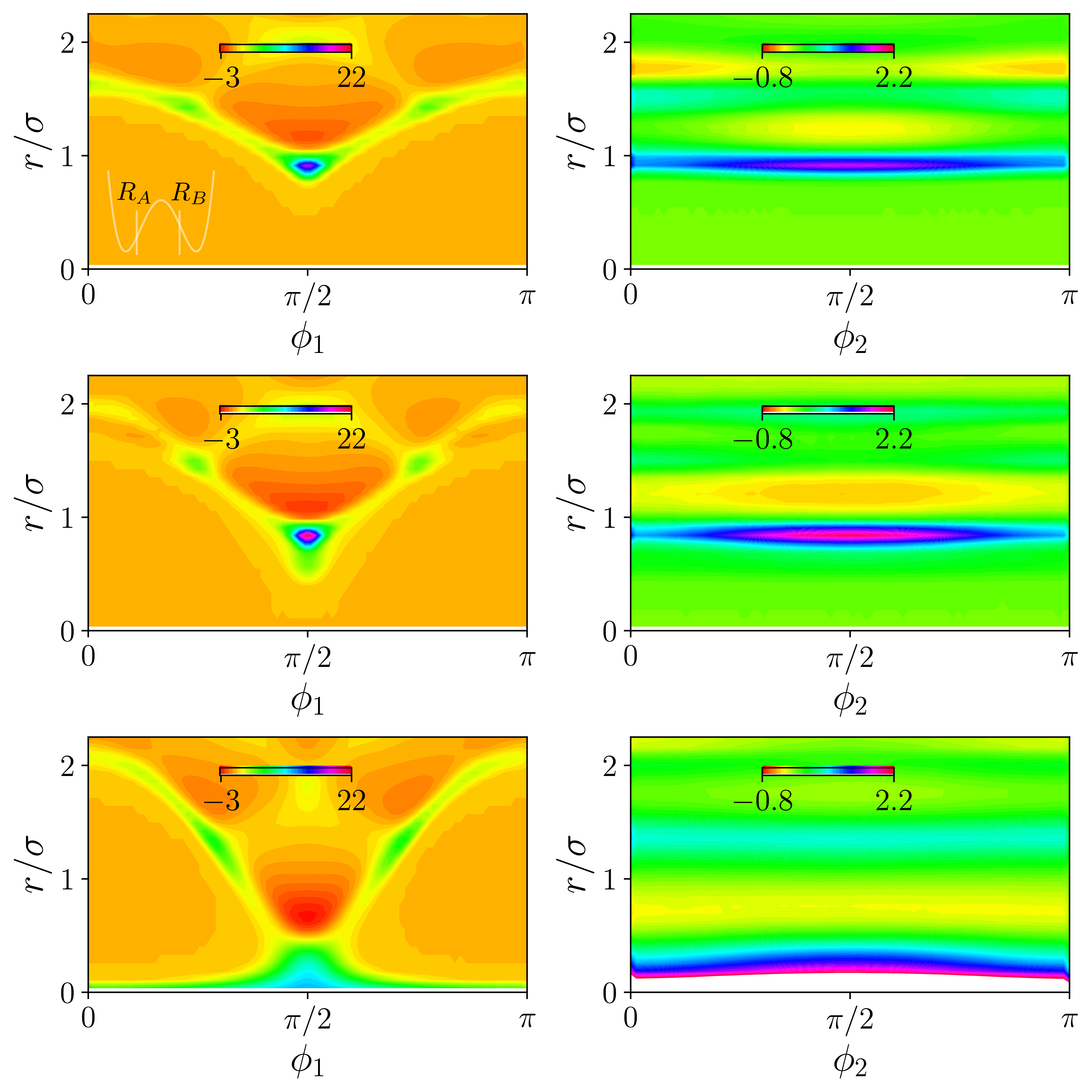}
\caption{Solvation structure of the dimer by the active bath.  Difference in pair distributions $\Delta g_X(r, \phi_{1})$ (Left) and  $\Delta g_X(r, \phi_{2})$ (Right). Configurations are conditioned such that the bath is sampled with the dimer  in the collapsed state $\Delta g_A(r, \phi_{1,2})$ (Top), the transition region $\Delta g_{AB}(r, \phi_{1,2})$ (Middle) or the extended state $\Delta g_B(r, \phi_{1,2})$ (Bottom).}
\label{sm3}
\end{figure}


\section{Nonequilibrium solvation structure}

We have studied how the solvation structure around the dimer evolves with activity. 
We compute the two-dimensional pair distribution functions for the position of the solvent around the dimer bond within a conditioned  steady state ensemble average 
\begin{equation}
g_X(r, \phi_1)  = \frac{1}{\rho_\mathrm{s} \rho_\mathrm{d}}  \frac{\langle \sum_{i \in \mathrm{s}} \delta (\mathbf{r}_\mathrm{cm}^\mathrm{d}) \delta (r-|\mathbf{r}_i-\mathbf{r}_\mathrm{cm}^\mathrm{d}|) \delta(\phi_1-\mathrm{arccos}(\mathbf{r}_i \cdot \mathbf{R} ) )h_X(R)  \rangle}{ \langle h_X(R) \rangle },
\end{equation}
with the center of the dimer bond $\mathbf{r}_\mathrm{cm}^\mathrm{d} = (\mathbf{r}_1+\mathbf{r}_2)/2$ as a reference. Here, $r$ is the radial distance between $\mathbf{r}_\mathrm{cm}^\mathrm{d} $ and surrounding bath particles, which make an angle $\phi_1$  with the bond vector $\mathbf{R} = \mathbf{r}_1-\mathbf{r}_2$. The indicator function $h_X(R)$ restricts configurations where the bond length $R$ falls into state $X$. Similarly, we probe the orientation of active solute particles around the dimer bond vector with the pair distribution 
\begin{equation}
g_X(r, \phi_2) =  \frac{1}{\rho_\mathrm{s} \rho_\mathrm{d}} \frac{ \langle \sum_{i \in \mathrm{s}} \delta (\mathbf{r}_\mathrm{cm}^\mathrm{d}) \delta (r-|\mathbf{r}_i-\mathbf{r}_\mathrm{cm}^\mathrm{d}|) \delta(\phi_1-\mathrm{arccos}(\mathbf{e}[\theta_i] \cdot \mathbf{R} ) )h_X(R)  \rangle } {\langle h_X(R) \rangle }.
\end{equation}
where $\phi_2$ is the angle between a bath director and the dimer bond, and $\rho_\mathrm{s} \sigma^2 = 0.6$ is the density of the solvent, and $\rho_\mathrm{d} \sigma^2 = 0.008$ is the  density of the dimers. To compute $g_X(r, \phi_i)$ we average over configurations sampled from $24$ simulations each with a length of $2 \times 10^8 $ time-steps.  

In Fig. \ref{sm3},  we consider the change in the pair distributions $\Delta g_X(r, \phi_{i = 1,2}) = g_X(r, \phi_{i = 1,2}, v_\mathrm{0} = 9)-g_X(r, \phi_{i = 1,2}, v_\mathrm{0} = 0)$ in an active bath with $v_\mathrm{0} = 9$ and its equilibrium counterpart at $v_\mathrm{0} = 0$. The different rows impose different conditions for the dimer bond distance $R = |\mathbf{R}|$ to be either primarily in state $X = A$ (top row), with  $R<  1.55 \sigma $, in the transition region $X = AB$ between states $1.55 \sigma <R< 1.65 \sigma$ (center row), or mostly in state $X = B$ (bottom row), with  $R> 1.65 \sigma$.

These results demonstrate that the rate enhancement is correlated with active particles dynamically wedging within the cross section of the dimer, pushing it apart into an extended state.
The left column of Fig. \ref{sm3} demonstrates that activity greatly enhances the packing of bath particles between the two bonded dimer particles, while the right column illustrates that bath particles preferentially orient perpendicular to the bond vector once far enough within the cross section. In state $A$, the active nature of the bath causes particles to push the dimer apart along $R$, as evidenced by the depletion for $\phi_2 = \pi/2$ and $r/\sigma > 1$ in the top-right panel of Fig. \ref{sm3}. The transition region, center-left, shows a significantly higher peak in the radial distribution function around $\phi_2 = \pi/2$, marking a decrease in the height of the effective free energy barrier along $R$. This analysis also illustrates the mechanism of increased stability in the active dimer extended state. Namely, Fig. \ref{sm3} bottom-left shows that the driven bath particles act to inhibit the extended state from closing. 

\bibliography{sm_rate.bib}